\documentclass[twocolumn]{aastex61}

\newcommand{\pcc}{\,{\rm cm}^{-3}}

\newcommand{\pc}{\, {\rm pc}}
\newcommand{\yr}{\, {\rm yr}}

\shorttitle{Chemical evolution of prestellar cores}
\shortauthors{Priestley et al.}

\begin{document}

\title{An efficient method for determining the chemical evolution of gravitationally collapsing prestellar cores}

\correspondingauthor{F. D. Priestley}
\email{fdp@star.ucl.ac.uk}

\author{F. D. Priestley}
\affiliation{Department of Physics and Astronomy, University College London, Gower Street, London WC1E 6BT, UK}

\author{S. Viti}
\affiliation{Department of Physics and Astronomy, University College London, Gower Street, London WC1E 6BT, UK}

\author{D. A. Williams}
\affiliation{Department of Physics and Astronomy, University College London, Gower Street, London WC1E 6BT, UK}



\begin{abstract}

We develop analytic approximations to the density evolution of prestellar cores, based on the results of hydrodynamical simulations. We use these approximations as input for a time-dependent gas-grain chemical code to investigate the effects of differing modes of collapse on the molecular abundances in the core. We confirm that our method can provide reasonable agreement with an exact numerical solution of both the hydrodynamics and chemistry while being significantly less computationally expensive, allowing a large grid of models varying multiple input parameters to be run. We present results using this method to illustrate how the chemistry is affected not only by the collapse model adopted, but also by the large number of unknown physical and chemical parameters. { Models which are initially gravitationally unstable predict similar abundances despite differing densities and collapse timescales, while ambipolar diffusion produces more extended inner depleted regions which are not seen in observations of prestellar cores. Molecular observations are capable of discriminating between modes of collapse despite the unknown values of various input parameters.} We also investigate the evolution of the ambipolar diffusion timescale for a range of collapse modes, metallicities and cosmic ray ionization rates, finding that it remains { comparable to or larger than} the collapse timescale during the initial stages for all models we consider, but { becomes smaller at later evolutionary stages. This confirms that ambipolar diffusion is an important process for diffuse gas, but becomes less significant as cores collapse to higher densities.}

\end{abstract}




\section{Introduction}
\label{sec:intro}
Dense ($\ge 10^4$ cm$^{-3}$), cold ($\sim 10$ K) regions in molecular clouds are the reservoir of gas for low mass star formation.
While the chemistry in these regions has been studied for many years and is broadly understood, the chemistry of star-forming cores remains controversial.
The distribution of core masses within the molecular
clouds is similar to the observed stellar initial mass function \citep{motte1998}, and young low mass
stars are frequently found to be associated with dense cores \citep{cohen1979}. The process by which a dense core evolves into a star can be partially deduced by observations of molecular line emission, and additionally by dust observations later on in the evolutionary process as the gas heats up. Nevertheless the very early stage of low mass star formation, when collapse begins, is 
elusive and there is no direct measurement of the mode(s) of collapse that the core undergoes. Of course, in the absence of any 
pressure force, a cloud of gas collapses in one free-fall time, $t_{ff}$ = $\sqrt{\frac{3\pi}{32G\rho}}$, but in reality, pressure gradients within a cloud will act to inhibit collapse. A common solution to the hydrostatic equlibrium equation is the well known Bonnor-Ebert (BE) sphere \citep{bonnor1956}, which describes the density profile of a pressure-confined, self-gravitating isothermal sphere in hydrostatic equilibrium. BE spheres are unstable to gravitational collapse if the ratio between central and outer densities exceeds a critical value, which leads to a critical mass for instability. \citet{kandori2005}
were able to match the density profiles of prestellar cores with BE spheres close
to this critical mass, while \citet{kirk2005} found that
for brighter cores, the observed density
profiles did not match critical BE spheres. This suggests that while cores may pass through a phase similar
to an unstable, thermally supported sphere, it is not sufficient to describe
their entire evolution.

In fact, a significant number of dense cores are observed to be rotating \citep{goodman1993}, and as angular momentum must be conserved during collapse,
the infalling gas would be expected to form a rotating disc rather than
collapsing directly into the centre. Simulations of the collapse of rotating gas clouds \citep{norman1980,matsumoto1997} find that although there is a complex structure of periodic shock wave formation at the centre,
the overall picture is of a disc forming in the plane of rotation, with
the central density and 
flatness increasing over time. While rotation does not seem to change the qualitative nature of the
collapse, it may be important in determining the subsequent evolution
of the central object. Rotation also causes the "angular momentum problem", which is
that, theoretically, rotating prestellar cores have much more angular momentum than a star could
contain without breaking up \citep{prentice1971}. Proposed solutions
include the fragmentation of the core into multiple protostellar objects
\citep{matsumoto2003} and the removal of angular momentum by
magnetic fields, known as magnetic braking \citep{basu1994}. Simulations of the fragmentation and collapse of a magnetised filament
\citep{nakamura1995,tomisaka1996} show that, as with
rotation, once a cloud has begun to collapse dynamically, magnetic forces are
not sufficient to halt it. However, their conclusions depend on the coupling between the magnetic field and the gas and this is determined by the fractional ionization, which itself is controlled by the chemical evolution during the collapse and not included in these works.

The density structure of prestellar cores can be inferred from observations
of either the dust continuum, or line emission from molecules. Both methods
require a conversion factor to relate either the dust emissivity or
the column density of the observed molecular species to the total gas density;
these conversion factors introduce systematic uncertainties. Additionally, at typical distances
observations of the central regions are limited by resolution, while in the
low-density outer regions the lower signal-to-noise is a further source of 
uncertainty. The differences in density structure between many proposed 
models of collapse are often not large enough for observations to 
conclusively favour one over the others. However,
the timescale for molecular abundances in a core to reach equilibrium
levels are comparable to the timescales involved in gravitational collapse,
which means that the chemical composition of a cloud should be sensitive
to the hydrodynamical situation. Simulations of the chemical evolution of
a core undergoing collapse \citep{rawlings1992,bergin1997,aikawa2001} have found that this is the case, with different hydrodynamical
models giving significant differences in the abundances of some
molecules. This raises the possibility that molecular abundances could be
used as an observational test of theories of star formation. This is our motivation.

However, there are difficulties in coupling detailed chemical models to a hydrodynamical
code, especially when magnetic fields are included, and simpler chemical
models cannot necessarily be relied on to give accurate molecular abundances. 
For example, \citet{aikawa2001} and \citet{rawlings1992} used analytical solutions for isothermal
collapse, found by \citet{larson1969} and \citet{shu1977} respectively, but these
are not necessarily applicable to real situations - the \citet{larson1969} solution only
agrees with simulations at small radii \citep{hunter1977}, while the \citet{shu1977} solution
begins from a static singular isothermal sphere, which is unlikely to be a realistic initial
state. \citet{lee2004} obtained the initial conditions for the \citet{shu1977} solution
using a series of progressively denser BE spheres evolving quasistatically, before continuing to
follow the chemical evolution throughout the subsequent dynamical collapse.
\citet{aikawa2005} improved on this earlier work and performed full hydrodynamical calculations of the collapse of BE spheres,
while \citet{li2002} used a simplified, one-dimensional model of magnetic
forces, to make the simulation computationally feasible. \citet{hincelin2013,hincelin2016} followed the chemical evolution of tracer particles 
in a fully 3-dimensional magnetohydrodynamical (MHD) simulation, 
assuming that the neutral gas is perfectly coupled to the magnetic field. \citet{tassis2012}
used the thin disc approximation to investigate the effects of non-ideal MHD, where 
the neutral and ionised parts of the gas behave differently. \citet{keto2015} used a simplified chemical
network to investigate the effects of different collapse modes on the line profiles of CO and H$_2$O, finding that
quasi-equilibrium contraction of a BE sphere was best able to reproduce observations.

In this paper we propose a different approach: we parametrize the results of hydrodynamical simulations
of collapsing prestellar cores to describe how the density behaves as a function of radius
and time for different models, and incorporate these parametrizations into a gas-grain time dependent chemical model. Although less accurate than a simultaneous solution of both the hydrodynamics and chemistry, this approach removes the need for simplifications in either area, while also being much less computationally expensive, and so enabling the exploration of larger regions of parameter space than has so far been feasible.

This paper is laid out as follows: in Section~\ref{sec:method}, we describe our parametrizations of hydrodynamical simulations, and we discuss the chemical model into which we incorporate them; in Section~\ref{sec:results} we present the results of our grid of models, showing how the abundances of key molecules are affected; in Section~\ref{sec:timescales} we investigate whether the inclusion of ambipolar diffusion affects star formation timescales; we briefly discuss our findings and conclude in Section~\ref{sec:discussion}. In Appendices \ref{sec:eqs} and \ref{sec:vr} we list the functions used in our parametrization of the numerical simulations.

\section{Methodology}
\label{sec:method}

\subsection{Parametrization of numerical simulations}

{ Empirical models were developed to reproduce the results from four numerical simulations of collapsing prestellar cores: \citet{aikawa2005} used a BE sphere as the initial configuration, with the density increased by a factor of either 1.1 (model BES1) or 4 (BES4) to take the core out of equilibrium and instigate collapse; \citet{nakamura1995} studied the collapse of a magnetically supported filament when a density perturbation is applied (MS); and \citet{fiedler1993} followed the evolution of a core as magnetic support is removed through ambipolar diffusion (AD).

{ Each of these studies produced the density profile (number density for BES1, BES4 and AD, mass density for MS) of the core as a function of time during the collapse. We extracted data from published plots of the density profiles at different times, as shown in Figure \ref{fig:bes1-fit} with data from \citet{aikawa2005} as an example.}

For each density profile, a function, 
depending on position in the core, 
 reproducing its shape was found. In the two models including magnetic effects (MS and AD), density profiles were given for both the radial ($z = 0$) and $z$ axes. Only the density profile in the radial direction was considered, as the core rapidly collapses into a thin disc so that structure along the $z$-axis is less significant. For the BES1, BES4 and AD models, a function of the form
\begin{equation}
n(r) = \frac{n_0}{1 + (\frac{r}{r_0})^a}
\label{eq:profile}
\end{equation}
with \(n_0\), \(r_0\) and \(a\) free parameters determining the central density, the width of the central density peak, and the slope of the profile, provided a good fit. \citet{tafalla2002} used Eq.~\ref{eq:profile} to fit the observed density profiles of prestellar cores. For the MS model, the radial profiles could not be approximated by Eq.~\ref{eq:profile}. The equilibrium density of the filament is given by an equation of the form
\begin{equation}
\rho(r) = \rho_0 \left(1 + \left(\frac{r}{r_0}\right)^2\right)^{-a}
\label{eq:profile2}
\end{equation}
which was adapted to reproduce the data by changing the { outer exponent $a$}, so that the slope at large \(r\) is the same as in the simulated data.

The values of the free parameters were chosen to approximate the density profiles at each time point given. Figure~\ref{fig:bes1-fit} shows the result for the \citet{aikawa2005} data. { For each parameter, the time evolution was also approximated by a similar method. For example, the central density parameter's time evolution was found to be well reproduced by}
\begin{equation}
  \log n_0(t) = A(t_0 - t)^{a} + B
\end{equation}
{ for all simulations, where $t_0$ is the simulation's duration. $a$ was chosen such that $\log n_0$ is approximately linear with $(t_0 - t)^a$, and the coefficients $A$ and $B$ can then by found by linear regression.} Figure~\ref{fig:bes1-rho0fit} shows the variation of the central density parameter, \(n_0\), with time for the data from \citet{aikawa2005}, along with the resulting approximation to the time dependence of this parameter. Once all the parameters have been approximated in this way, the density can be calculated as a function of time and space, shown in Figure~\ref{fig:bes1-final} compared with the original data. Figures \ref{fig:bes4-final}, \ref{fig:ms-final} and \ref{fig:ad-final} show the corresponding density profile approximations for the BES4, MS and AD collapses respectively. The equations used to approximate the time dependence of the density profiles are given in Appendices \ref{sec:eqs} and \ref{sec:vr}. { The maximum discrepancies between the simulations and approximated densities are 34\%, 240\%, 66\% and 53\% for the BES1, BES4, MS and AD cases respectively, while the average discrepancies are 10\%, 26\%, 16\% and 16\%. The large ($>100 \%$) errors in the BES4 approximation occur only at late times and large radii - otherwise the agreement with the data is at a similar level to the other collapses.}

{ Our approximations give the time evolution of the density at a given radius. However, during collapse the individual parcels of gas do not remain at a constant radius, but move inwards, leading to a different density evolution than for fixed $r$. For the BES1 and BES4 models, we determine the new radius of a parcel at each time step by calculating the mass interior to its initial radius at $t = 0$, $M(<r_0)$, and finding the radius at the given time which encloses the same mass. The MS and AD approximations are not spherically symmetric, so this approach would require knowledge of the density at each point. Instead, we use the same methods as for the density to approximate the radial velocity profiles, and use these to calculate the new parcel radius at each timestep. The gas density versus time, for parcels of differing initial radii, in the BES1 collapse is shown in Figure \ref{fig:traj}.}

\begin{figure}
\centering
\plotone{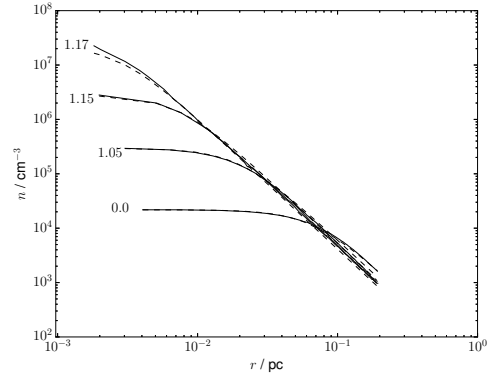}
\caption{Density profiles taken from \citet{aikawa2005} (solid lines), with the approximate profiles calculated using Eq.~\ref{eq:profile} (dashed lines). The labels indicate the time since collapse in \(10^6 \, \mathrm{yr}\).}
\label{fig:bes1-fit}
\end{figure}

\begin{figure}
\centering
\plotone{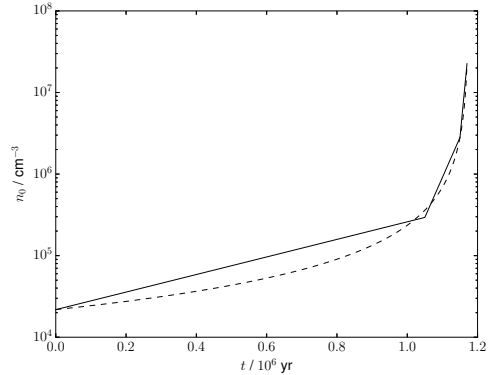}
\caption{Central density \(n_0\) against \(t\) for the approximations to the \citet{aikawa2005} data (solid), with the approximate fit to the time evolution (dashed).}
\label{fig:bes1-rho0fit}
\end{figure}

\begin{figure}
\centering
\plotone{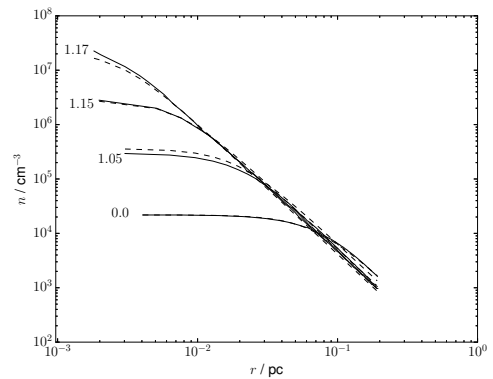}
\caption{As Figure~\ref{fig:bes1-fit}, but with the parameters for Eq.~\ref{eq:profile} calculated as a function of time.}
\label{fig:bes1-final}
\end{figure}

\begin{figure}
\centering
\plotone{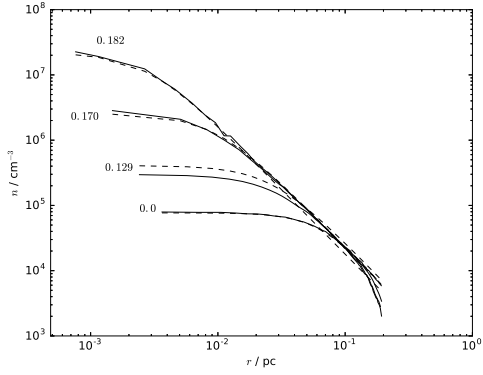}
\caption{As Figure~\ref{fig:bes1-final}, for the BES4 collapse}
\label{fig:bes4-final}
\end{figure}

\begin{figure}
\centering
\plotone{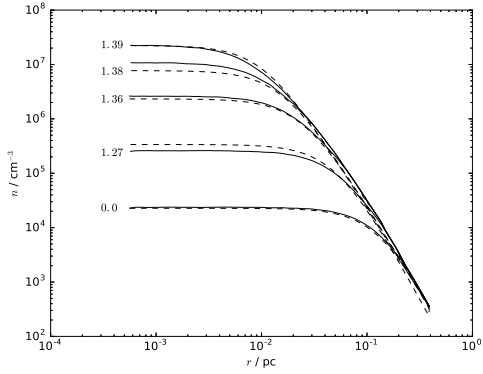}
\caption{As Figure~\ref{fig:bes1-final}, for the MS collapse}
\label{fig:ms-final}
\end{figure}

\begin{figure}
\centering
\plotone{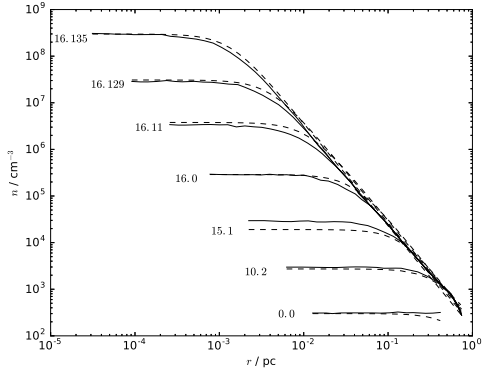}
\caption{As Figure~\ref{fig:bes1-final}, for the AD collapse}
\label{fig:ad-final}
\end{figure}

\begin{figure}
  \centering
  \plotone{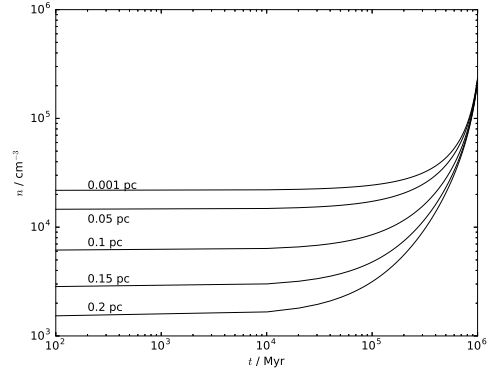}
  \caption{Parcel density versus time for the BES1 collapse, for different initial parcel radii.}
  \label{fig:traj}
\end{figure}

\subsection{Chemical modelling}

The four density approximations were used as input for the chemical code UCL\_CHEM. The code is described in \citet{viti2004} and references therein. This code has since been made public (https://uclchem.github.io/) and is fully explained in \citet{holdship2017}. Here we briefly summarize its characteristics: UCL\_CHEM is a time dependent gas-grain chemical model that calculates the abundances of atoms and molecules in the gas and dust in the interstellar medium as a function of time under chemical and physical conditions set by the user. The original version of the code uses free-fall collapse to determine the density from the diffuse state to the final density of the gas where the star is born. Initial atomic elemental abundances are provided to UCL\_CHEM which then self-consistently calculates  gas-phase chemistry, as well as sticking on to dust particles with subsequent surface processing. For the reaction network we used the UMIST 2012 network \citep{mcelroy2013}, and freeze-out and grain surface reactions as described in \citet{holdship2017}.

We chemically model our four parametrized numerical simulations: the collapse of an unstable (BES1) or highly unstable (BES4) Bonnor-Ebert sphere, collapse against magnetic support (MS) and collapse resulting from ambipolar diffusion (AD). A grid of models was run for each case to investigate the effects of changing other input parameters: the cosmic ray ionisation rate \(\zeta\) (in units of $\zeta_0 = 1.3 \times 10^{-17}$ s$^{-1}$), metallicity $Z$, and { the desorption efficiency parameters \(\epsilon\), \(\phi\) and \(Y_{\rm UV}\), corresponding to the number of molecules desorbed per H$_2$ molecule formed, per cosmic ray impact and per UV photon (produced by cosmic rays) absorbed respectively \citep{roberts2007}. For our fiducial desorption efficiencies, H$_2$ formation is the dominant desorption mechanism, as found by \citet{roberts2007}.} The values adopted for each model are given in Table~\ref{tab:grid}. Each model was run once for each of the density approximations. We assume an external radiation field of 1 Habing, and an external extinction at the core boundary of 3 mag, the value used by \citet{aikawa2005}. The extinction from the core itself is calculated by integrating the density profile to the boundary, ($0.2 \pc$ for the BES1 and BES4 approximations, $0.5 \, {\rm pc}$ for MS and $0.75 \pc$ for the AD case) before the onset of collapse, { giving maximum extinctions (at the centremost parcel) of 6.4 (BES1), 15.2 (BES4), 7.8 (MS) and 3.3 (AD) mag}. { We used 13 gas parcels (14 for AD) at initial radii spaced to cover the entire range of the cores, but with an emphasis on the more rapidly evolving central regions.}

The initial central number densities of the models are \(n = 2.2 \times 10^4 \: \mathrm{cm}^{-3}\) (BES1 and MS), \(8 \times 10^4 \: \mathrm{cm}^{-3}\) (BES4) and \(300 \: \mathrm{cm}^{-3}\) (AD). The MS equations are in terms of dimensionless variables, which can be converted into physical values by choosing the intial central density, \(\rho_c\), and the isothermal sound speed, \(c_s^2 = \frac{kT}{\mu m_H}\). The gas was assumed to be at a temperature of \(10 \, \mathrm{K}\) and composed entirely of molecular hydrogen for the purposes of calculating the mean molecular mass, giving \(c_s = 203 \, {\rm m s}^{-1}\), and we set $\rho_c = 3.67 \times 10^{-20} \, {\rm g cm}^{-3}$ to give an initial central number density equal to the BES1 model. The initial number density at each point is then given by the relevant equation for the density profile, and we allow the chemistry to evolve for 1 Myr before the onset of collapse at this density. The models were run until the central density reached \(10^8 \: \mathrm{cm}^{-3}\). { The elemental abundances relative to H for our standard model, the solar values given by \citet{asplund2009}, are listed in Table \ref{tab:abun} - the abundances for elements other than H and He in models with varying metallicity are multiplied by the value of $Z$.}

\begin{table}
\centering
\caption{Model input parameters}
\label{tab:grid}
\begin{tabular}{cccccc}
\hline
\hline
Model & \(\zeta\)/$\zeta_0$ & Z/Z$_{\odot}$ & \(\epsilon\) & \(\phi\) & \(Y_{\rm UV}\) \\ \hline
A & 1 & 1 &  0.01 & \(10^5\) & 0.1 \\ \hline
B1 & 5 & 1 &  0.01 & \(10^5\) & 0.1 \\ \hline
B2 & 10 & 1 &  0.01 & \(10^5\) & 0.1 \\ \hline
C1 & 1 & 0.3 &  0.01 & \(10^5\) & 0.1 \\ \hline
C2 & 1 & 1.5 &  0.01 & \(10^5\) & 0.1 \\ \hline
D1 & 1 & 1 &  0.1 & \(10^5\) & 0.1 \\ \hline
D2 & 1 & 1 &  1.0 & \(10^5\) & 0.1 \\ \hline
E1 & 1 & 1 &  0.01 & \(10^4\) & 0.1 \\ \hline
E2 & 1 & 1 &  0.01 & \(10^6\) & 0.1 \\ \hline
F1 & 1 & 1 &  0.01 & \(10^5\) & 0.001 \\ \hline
F2 & 1 & 1 &  0.01 & \(10^5\) & 1.0 \\ \hline
\end{tabular}
\end{table}

\begin{table}
  \centering
  \caption{Elemental abundances}
  \label{tab:abun}
  \begin{tabular}{cccc}
    \hline
    \hline
    Element & Abundance & Element & Abundance \\ \hline
    H & $1.0$ & N & $6.8 \times 10^{-5}$ \\ \hline
    He & $0.085$ & S & $1.3 \times 10^{-5}$ \\ \hline
    C & $2.7 \times 10^{-4}$ & Si & $3.2 \times 10^{-5}$ \\ \hline
    O & $4.9 \times 10^{-4}$ & Cl & $3.2 \times 10^{-7}$ \\ \hline
  \end{tabular}
\end{table}

\section{Results}
\label{sec:results}

\subsection{Molecular abundances across the different modes of collapse}

{ Figure \ref{fig:n2e5} shows the density profiles of the four collapse modes at the point when the central number density, $n_0$, reaches $2 \times 10^{5} \pcc$. The BES1 and MS profiles decrease more rapidly with distance than for the BES4 and AD approximations, which have similar densities up to the end of the BES4 core at $0.2 \pc$. However, the time taken to reach this point is much shorter for the BES4 case ($\sim 10^5 \yr$) than for AD ($\sim 10^7 \yr$), so the chemical evolution differs significantly despite the gas density being the same. The BES1 and MS modes both reach $n_0 = 2 \times 10^{5} \pcc$ after $\sim 10^6 \yr$, and as such the chemical evolution is similar. The densities at large ($\gtrsim 0.2 \pc$) radii for BES4/AD are much higher than for BES1/MS, due to the more rapid collapse of the outer parts of the core in the BES4 case, and the magnetic support of the outer regions for AD.

Figure \ref{fig:abunA} shows the abundances of four molecules versus radii for model A for the four density approximations, at a central density $n_0 = 2 \times 10^{5} \pcc$. In all cases, the CO abundance increases from a central minimum, where freeze-out has depleted most of the molecule onto grain surfaces, to a maximum value of $\sim 10^{-4}$. The BES1, BES4 and MS approximations all behave similarly in reaching the maximum, although for BES4 and MS the radius at which this occurs is larger due to the higher gas densities increasing the effect of freeze-out. The CO abundance in the AD collapse increases much more slowly, only reaching $10^{-4}$ at the edge of the core, whereas in the MS case the abundance begins to fall again towards the edge. This is due to the higher densities in the outer regions for the AD collapse, as magnetic support can still prevent material here from collapsing, unlike in the other three cases. The effect of the longer collapse duration is also apparent, as the AD abundances are far lower than the BES4 ones at comparable radii, despite the densities being very similar. The NH$_3$ abundance also increases from the centre to a maximum, before falling with radius in all models. The decline is much more gradual for the AD case than the other collapse modes, leading to an order of magnitude difference with the MS model by $r = 0.3 \pc$. HCO$^+$ shows similar behaviour, while for HCN the AD collapse mode produces a nearly constant abundance after reaching a value of $\sim 10^{-8}$, in contrast to the others.

\begin{figure}
\centering
\plotone{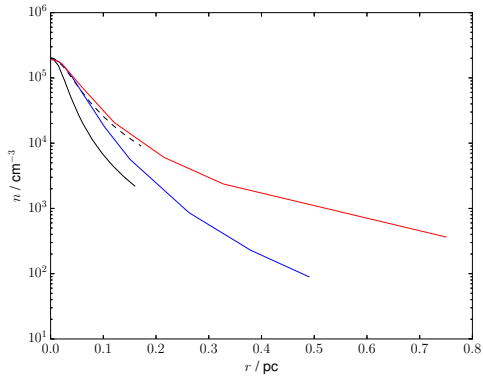}
\caption{Density profiles of the BES1 (solid black), BES4 (dashed black), MS (blue) and AD (red) approximations, at a central number density of $n_0 = 2 \times 10^{5} \pcc$.}
\label{fig:n2e5}
\end{figure}

\begin{figure*}
  \gridline{\fig{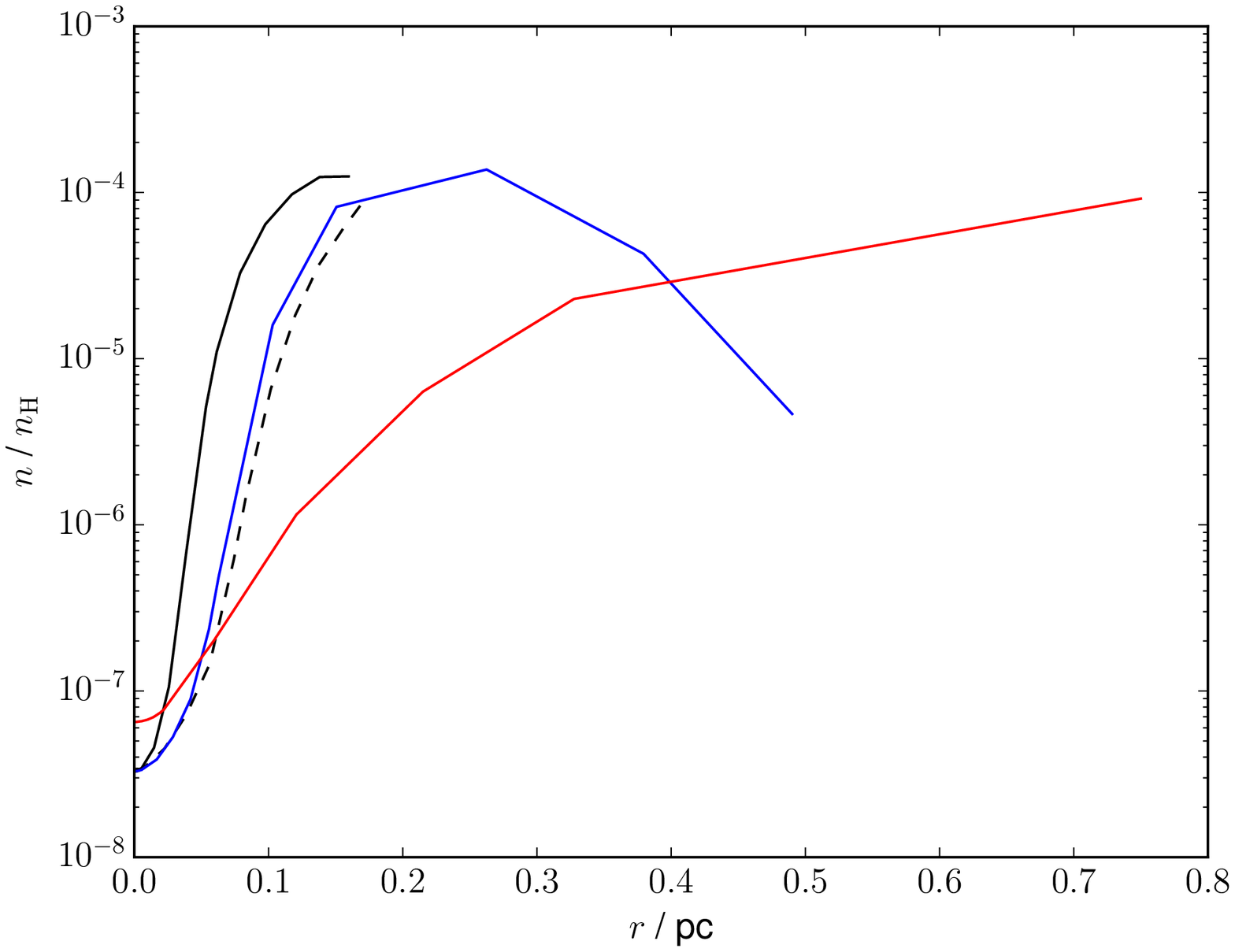}{0.5\textwidth}{(a) CO} \fig{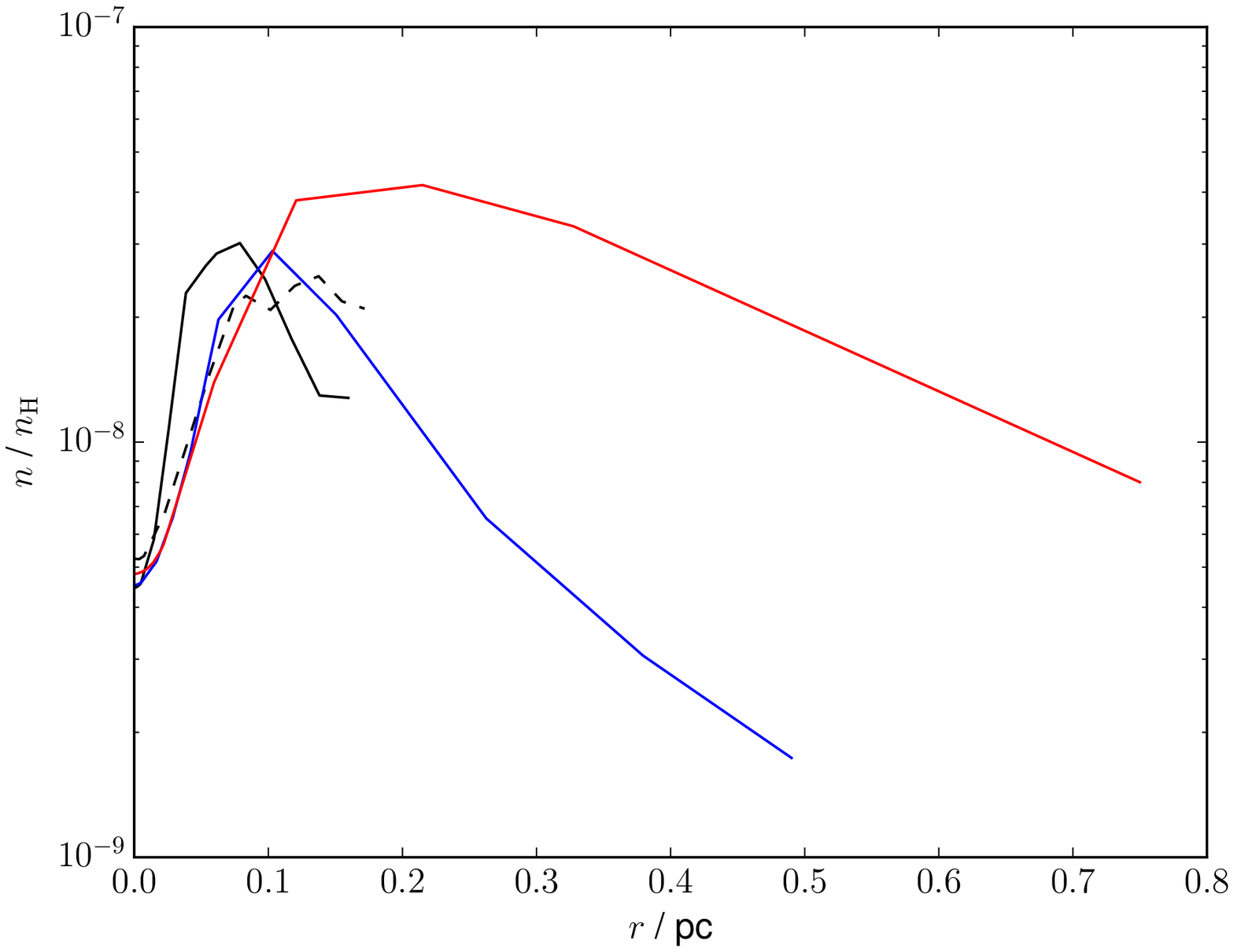}{0.5\textwidth}{(b) NH$_3$}}
  \gridline{\fig{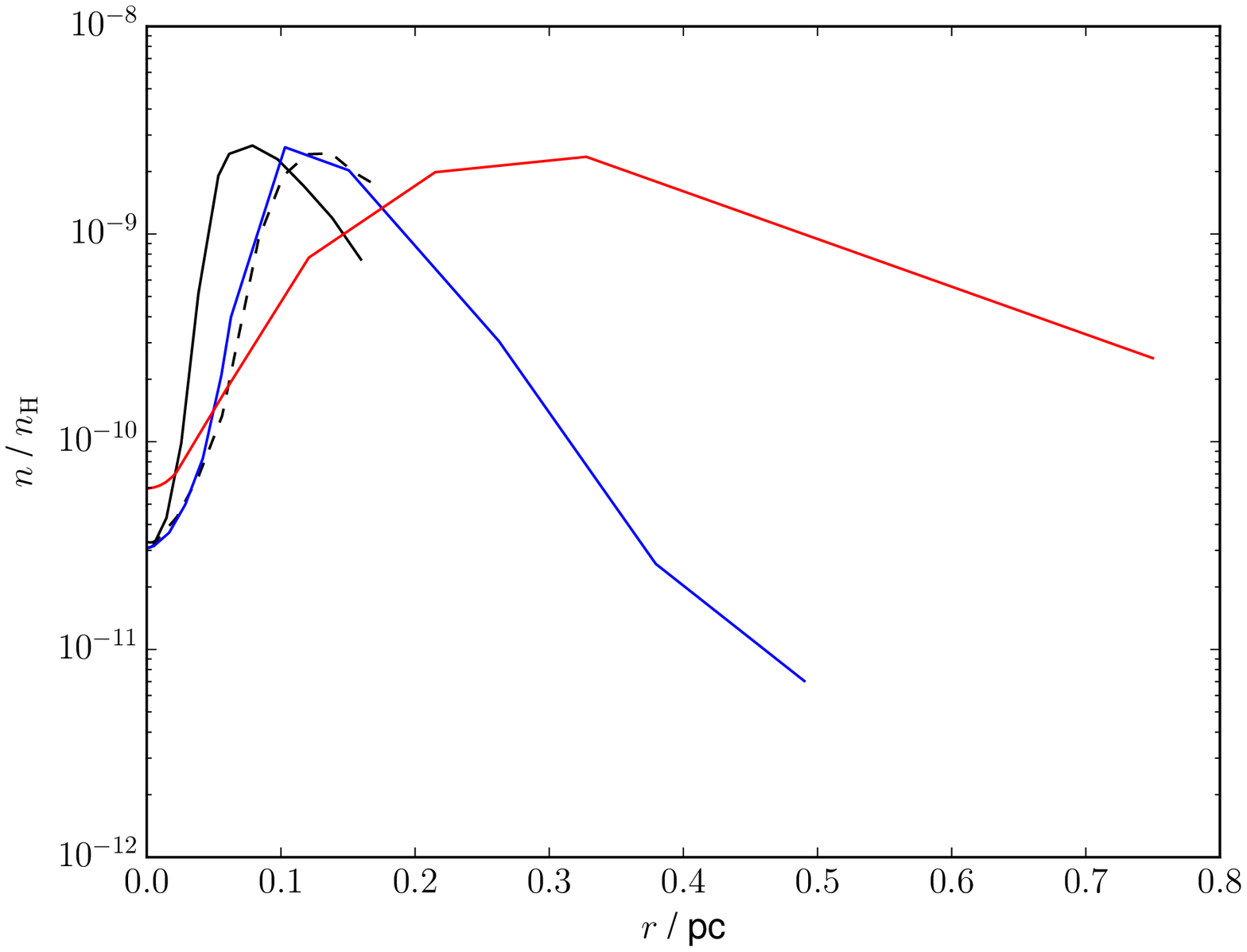}{0.5\textwidth}{(c) HCO$^+$} \fig{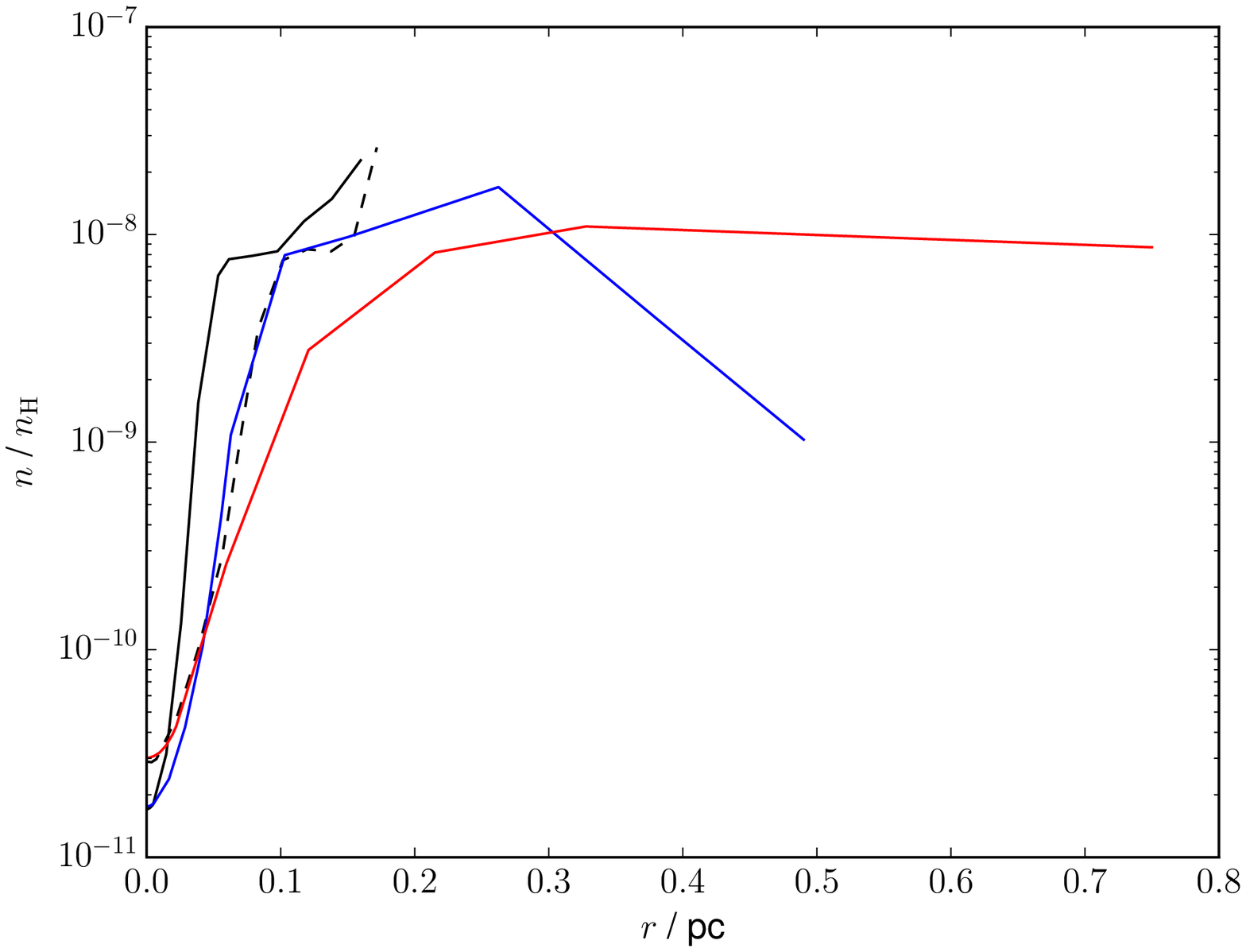}{0.5\textwidth}{(d) HCN}}
  \caption{Abundances of CO, NH$_3$, HCO$^+$ and HCN at a central density $n_0 = 2 \times 10^{5} \pcc$ for model A, using the BES1 (solid black), BES4 (dashed black), MS (blue) and AD (red) density approximations.}
  \label{fig:abunA}
\end{figure*}

\subsection{Cosmic ray ionization rate}

Raising the cosmic ray ionization rate increases the abundances of molecules in the centre of the core, regardless of the collapse mode, as the rate of desorption from grain surfaces is increased. Figure \ref{fig:coB} shows the CO abundances for models A, B1 and B2, for the BES1 and AD density approximations, at a central density $n_0 = 2 \times 10^{5} \pcc$. For the BES1 collapse, the abundance at larger radii is mostly unaffected, whereas for AD the A and B2 models show noticeably different behaviour, with the CO abundance an order of magnitude lower at the edge of the core for the B2 model due to the higher cosmic ray dissociation rate. Figure \ref{fig:hco+B} shows the HCO$^+$ abundance for the same models. The central abundances are again enhanced for models with higher ionization rates, but whereas for the AD collapse mode the abundance decreases towards the edge as with CO, for BES1 the abundance is higher throughout the cloud.

\begin{figure*}
  \gridline{\fig{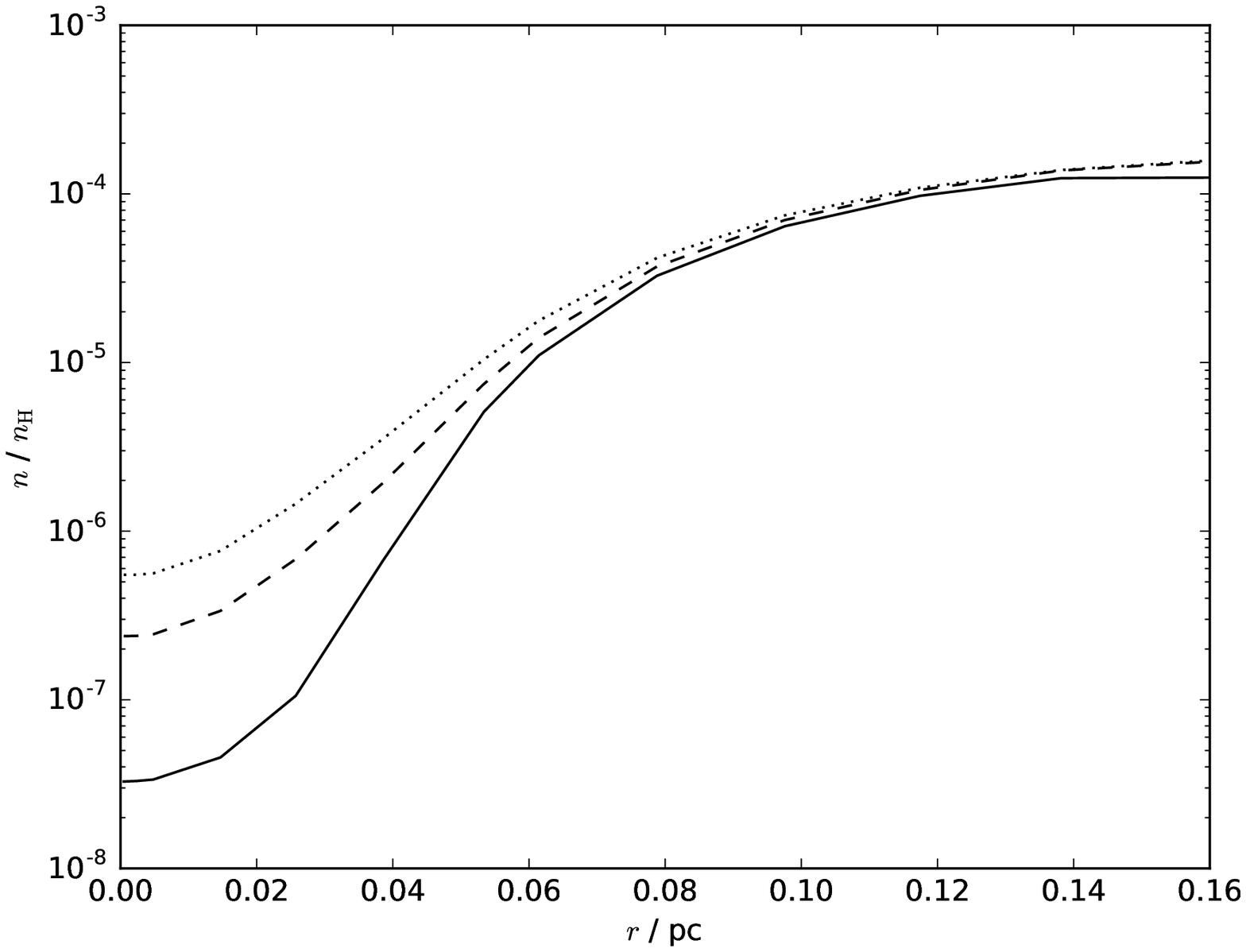}{0.5\textwidth}{(a) BES1} \fig{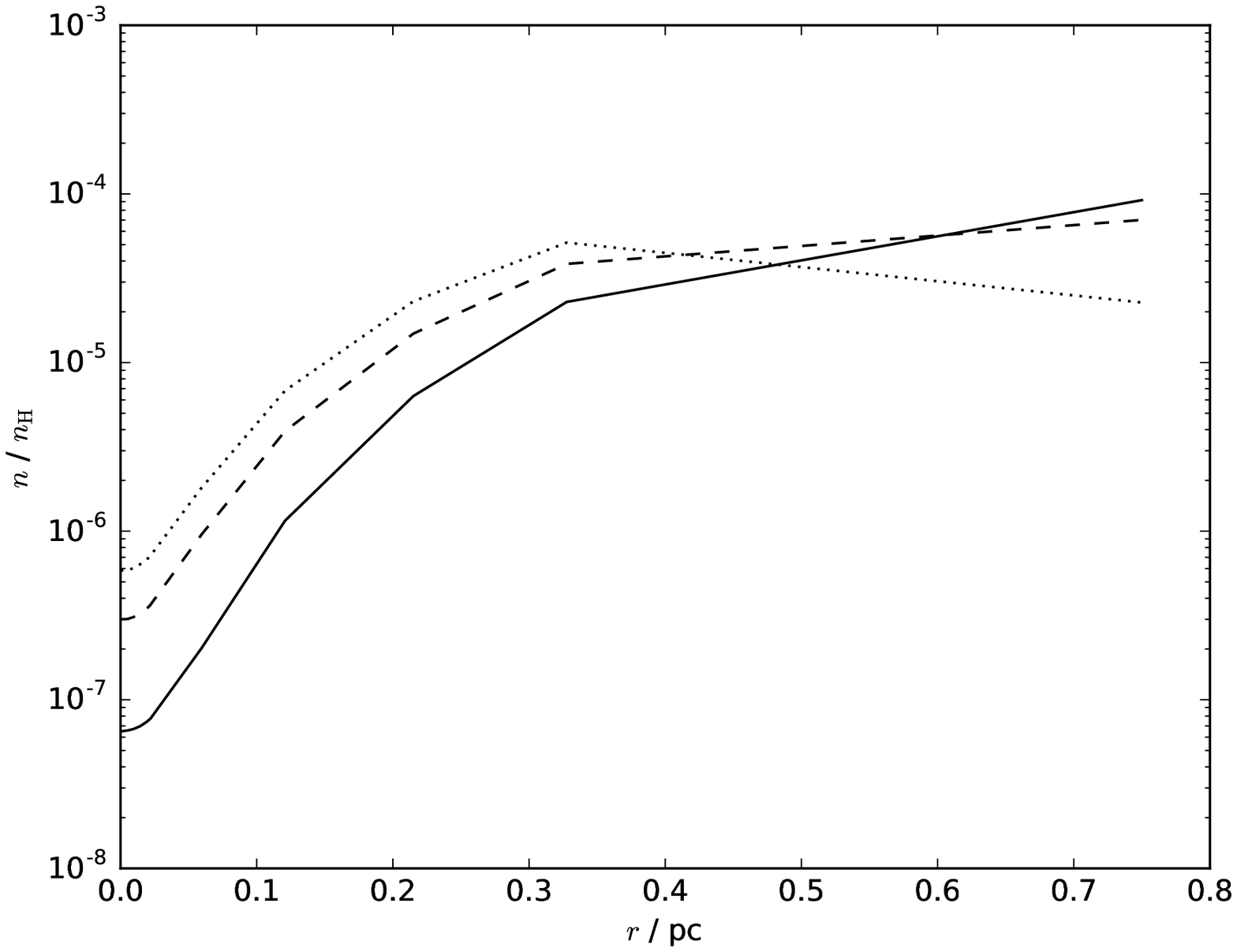}{0.5\textwidth}{(b) AD}}
  \caption{Abundance of CO versus radius at a central density $n_0 = 2 \times 10^{5} \pcc$ for models A (solid line), B1 (dashed line) and B2 (dotted line), using the BES1 (left) and AD (right) density approximations.}
  \label{fig:coB}
\end{figure*}

\begin{figure*}
  \gridline{\fig{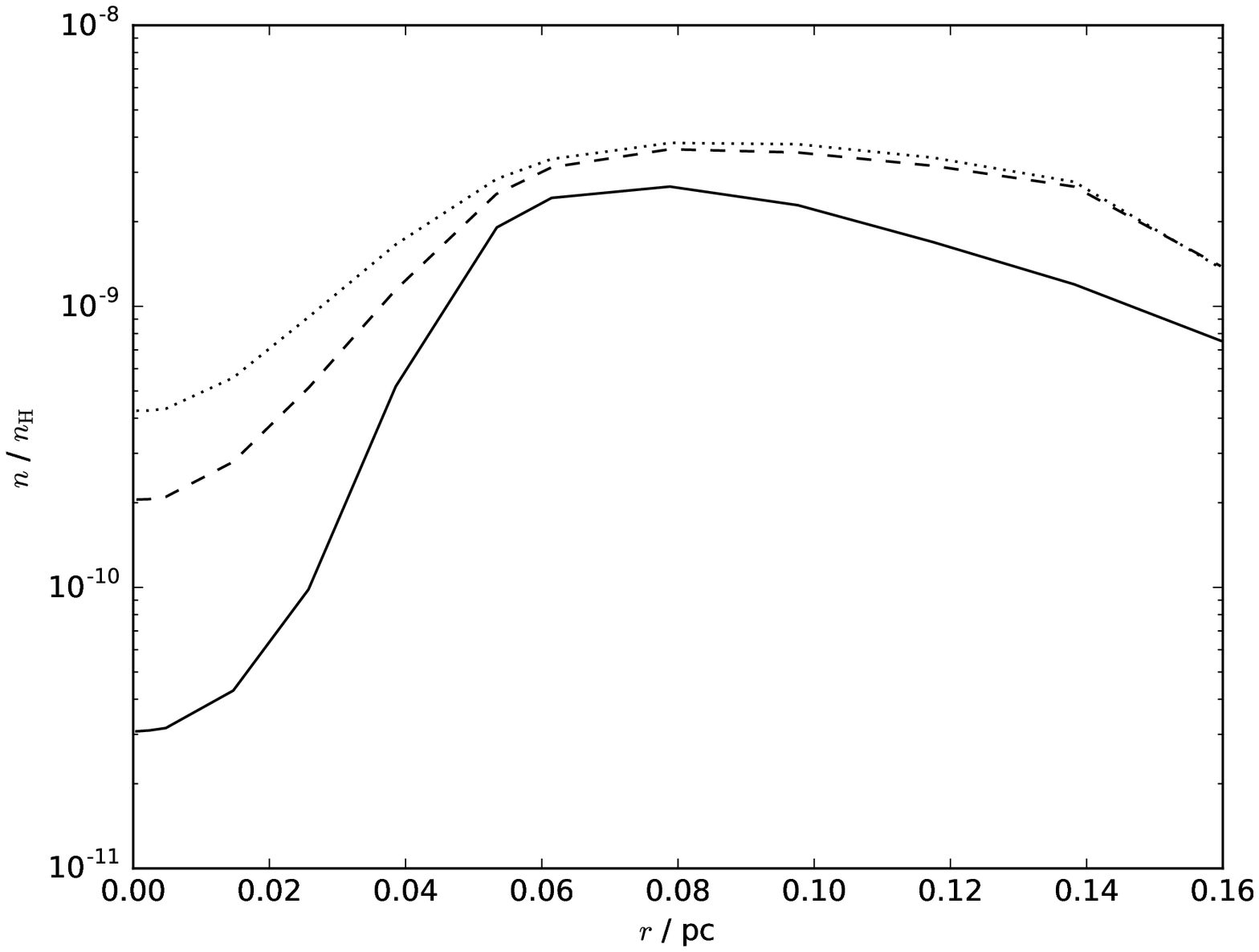}{0.5\textwidth}{(a) BES1} \fig{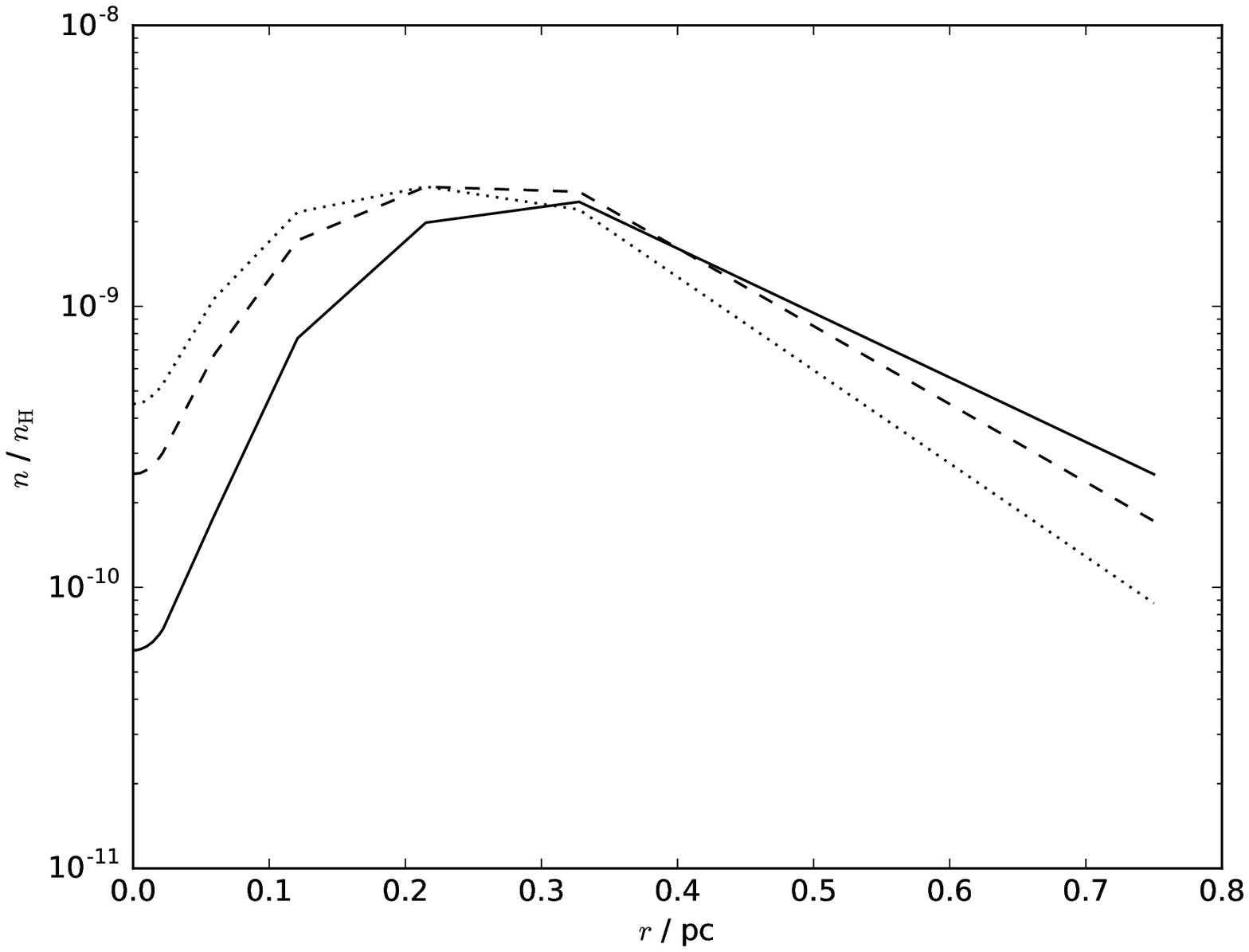}{0.5\textwidth}{(b) AD}}
  \caption{Abundance of HCO$^+$ versus radius at a central density $n_0 = 2 \times 10^{5} \pcc$ for models A (solid line), B1 (dashed line) and B2 (dotted line), using the BES1 (left) and AD (right) density approximations.}
  \label{fig:hco+B}
\end{figure*}

\subsection{Metallicity}

Changing the metallicity of the core usually results in a corresponding change in the molecular abundances, due to the different availability of atoms to form the molecules. However, some molecules are much less affected than others. Figure \ref{fig:bes1C} shows the abundances of CO and HCN for the BES1 collapse at a central density $n_0 = 2 \times 10^{5} \pcc$, for models A, C1 and C2. Whereas the CO abundance scales nearly linearly with the metallicity, the HCN abundance is virtually unchanged between models. { The main formation and destruction reactions for HCN both involve H$^+$, for which the abundance increases with decreasing metallicity (and vice versa), at least partially counteracting the effect of the changing elemental abundances on the HCN abundance.}

\begin{figure*}
  \gridline{\fig{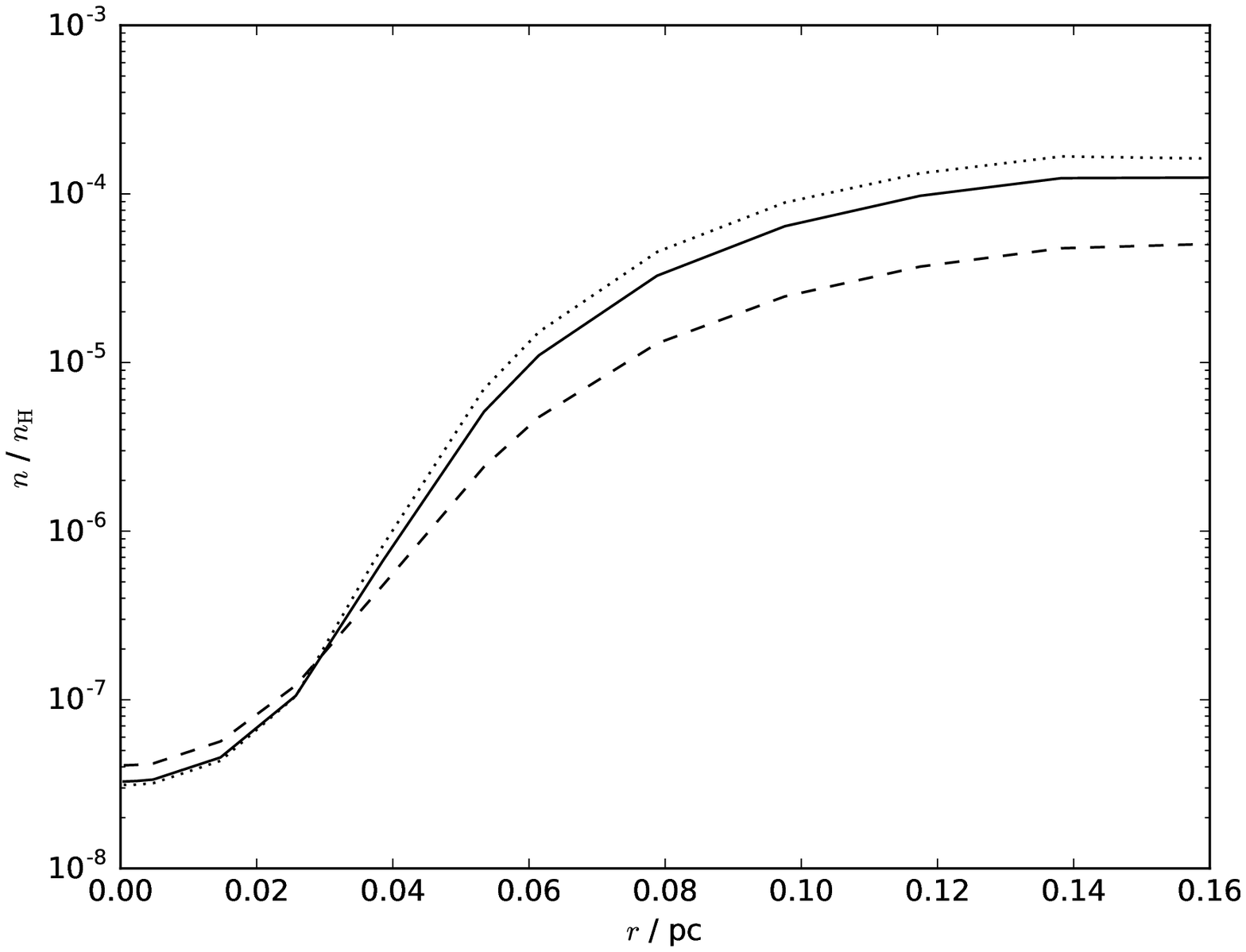}{0.5\textwidth}{(a) CO} \fig{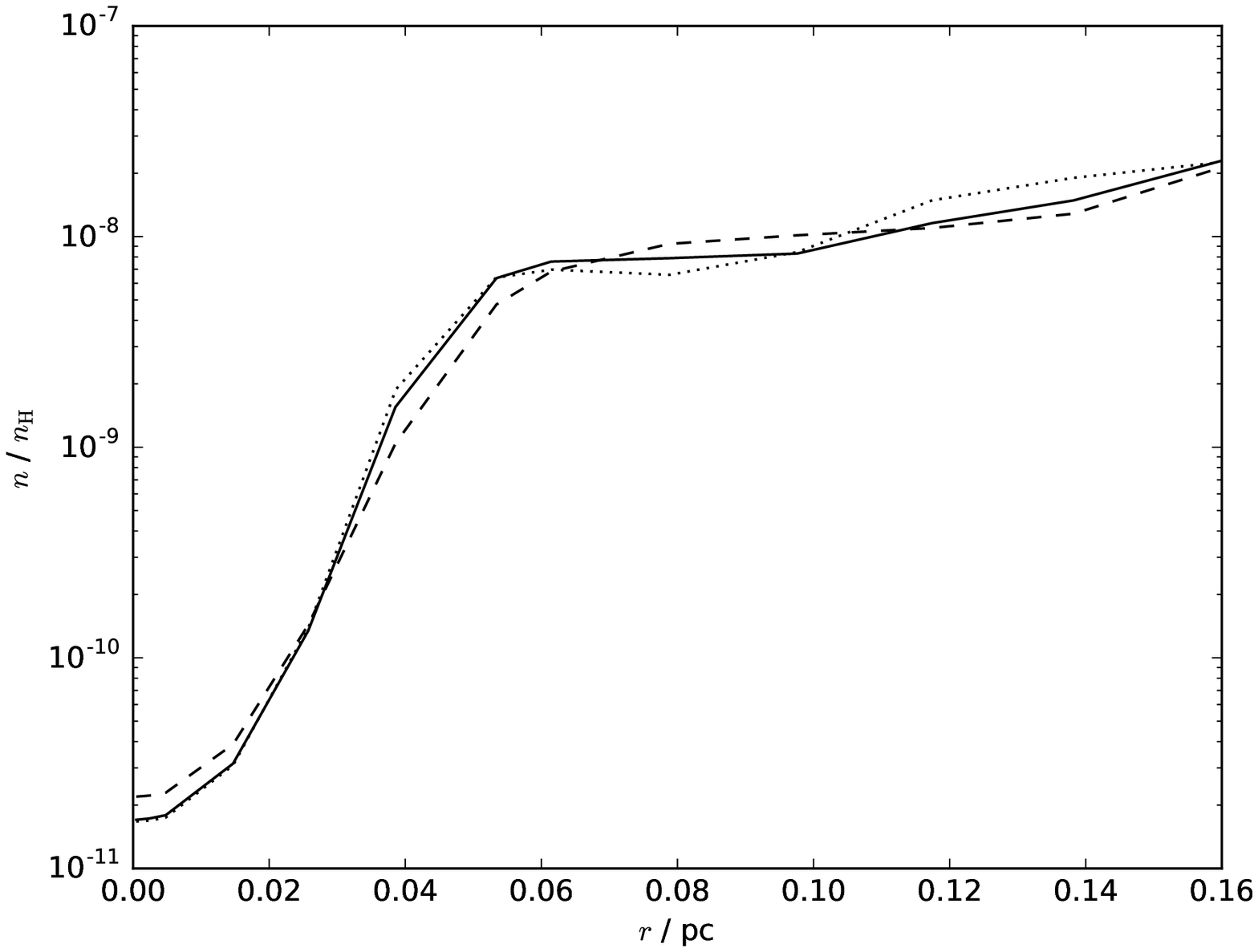}{0.5\textwidth}{(b) HCN}}
  \caption{Abundance of CO (left) and HCN (right) versus radius at a central density $n_0 = 2 \times 10^{5} \pcc$ for models A (solid line), C1 (dashed line) and C2 (dotted line), using the BES1 density approximation.}
  \label{fig:bes1C}
\end{figure*}

\subsection{Desorption efficiencies}

As with the cosmic ray ionization rate, increasing the desorption efficiences increases the molecular abundances, particularly in the denser central regions where more freeze-out has taken place. Figure \ref{fig:coD} shows the CO abundances for models A, D1 and D2, for the MS and AD density approximations, where the H$_2$ formation desorption efficiency $\epsilon$ has been modified. While the abundances in the central regions are affected similarly for both collapse modes, at larger radii the effect is negligible for the MS collapse, whereas the CO abundance reaches $10^{-4}$ much more rapidly for AD - for model D2, the abundance profile looks much more similar to the other collapse modes than for model A. { The BES1 and BES4 collapses behave similarly to MS for varying desorption efficiency, with very little change in the CO abundance beyond $0.1 \pc$.} The cosmic ray heating desorption efficiency $\phi$ has very little effect on the abundance of any molecule, despite a factor of $100$ difference between models E1 and E2. The cosmic ray induced photodesorption efficiency $Y_{\rm UV}$, however, does affect molecular abundances, in particular having a significant effect on the abundance of NH$_3$, which is not greatly affected by variation of the other parameters investigated.}

\begin{figure*}
  \gridline{\fig{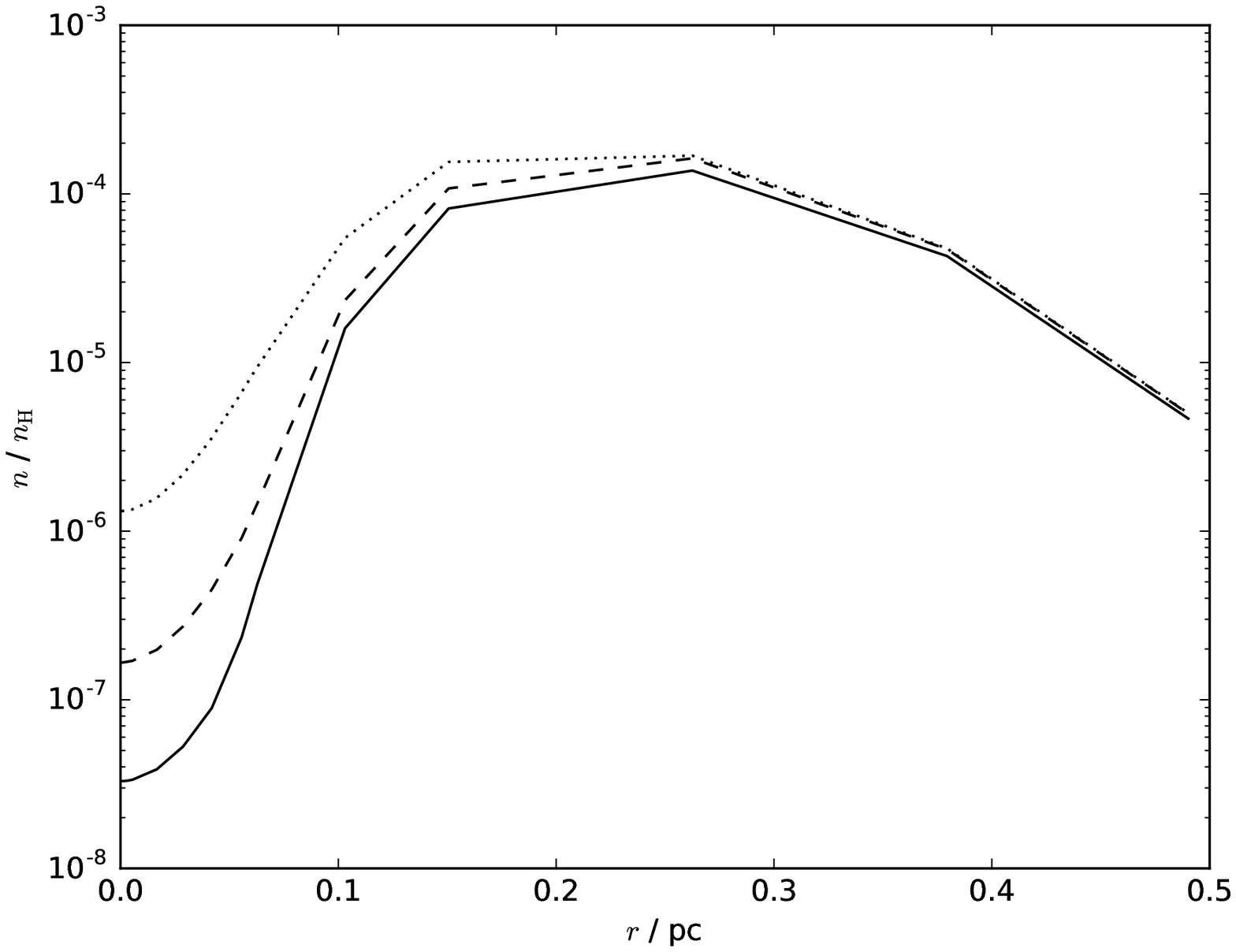}{0.5\textwidth}{(a) MS} \fig{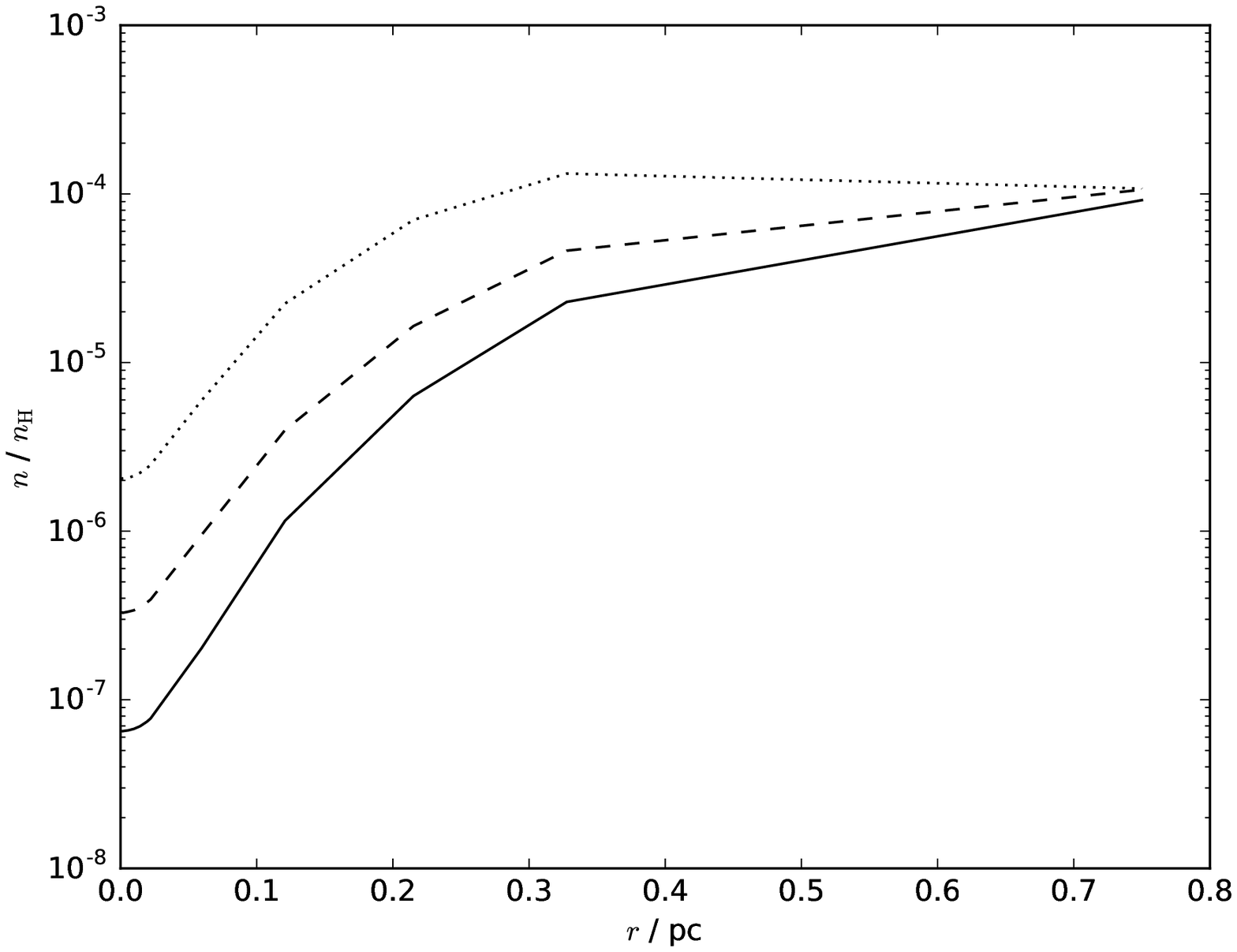}{0.5\textwidth}{(b) AD}}
  \caption{Abundance of CO versus radius at a central density $n_0 = 2 \times 10^{5} \pcc$ for models A (solid line), D1 (dashed line) and D2 (dotted line), using the MS (left) and AD (right) density approximations.}
  \label{fig:coD}
\end{figure*}

\section{Star formation efficiencies}
\label{sec:timescales}
Only one of our modes of collapse, AD, includes the effect of ambipolar diffusion. However, all prestellar cores are expected to be magnetised, and therefore ambipolar diffusion could be important for all collapse models. An estimate of the timescale on which ambipolar diffusion occurs is given by
\begin{equation}
  t_{amb} = 4 \times 10^5 (x_i/10^{-8}) yr
  \label{eq:tamb}
\end{equation}
\citep{mouschovias1979,hartquist1989}. If $t_{amb}$ is smaller than the free-fall timescale, magnetic pressure is unlikely to impede gravitational collapse, while if it is larger the impeding effects may be significant.

\citet{banerji2009} showed that the ambipolar diffusion timescale becomes very large as the fractional ionization increases and the magnetic field is strongly coupled to the collapsing core, which, in some cases, may halt the collapse and hence the formation of the star. We calculated the ambipolar diffusion timescale { at the centre of the core} using Eq.~\ref{eq:tamb} at { the beginning of the collapse, and at central densities of $10^6$ and $10^8 \pcc$, for the BES1, BES4 and MS approximations, for differing metallicities and cosmic ray ionization rates}, { using ionization fractions calculated in our chemical simulations}. We compared { these timescales} with the time it takes the gas to reach the final density ($10^8$ cm$^{-3}$). Table~\ref{tab:timescales} shows the collapse duration and $t_{amb}$ for this grid of models.

\begin{table*}
\centering
\caption{Collapse duration and ambipolar diffusion timescales at increasing density for BES1, BES4 and MS models with varying $\zeta$ and Z.}
\label{tab:timescales}
\begin{tabular}{ccccccc}
\hline
\hline
& & & & & $t_{amb}$/$10^6$ yr & \\ \cline{5-7}
Model & Z/Z$_{\odot}$ & $\zeta$/$\zeta_0$ & $t_{collapse}$/$10^6$ yr & Initial & $n_{\mathrm{H}} = 10^6$ cm$^{-3}$ & $n_{\mathrm{H}} = 10^8$ cm$^{-3}$ \\
\hline
BES1 A & $1.0$ & $1.0$ & $1.173$ & $0.80$ & $0.09$ & $0.08$ \\
BES1 B1 & $1.0$ & $5.0$ & $1.173$ & $2.51$ & $0.14$ & $0.09$ \\
BES1 B2 & $1.0$ & $10.0$ & $1.173$ & $4.70$ & $0.19$ & $0.10$ \\
BES1 C1 & $0.3$ & $1.0$ & $1.173$ & $0.62$ & $0.09$ & $0.08$ \\
BES1 C2 & $1.5$ & $1.0$ & $1.173$ & $0.95$ & $0.09$ & $0.07$ \\
BES4 A & $1.0$ & $1.0$ & $0.184$ & $0.22$ & $0.08$ & $0.07$ \\
BES4 B1 & $1.0$ & $5.0$ & $0.184$ & $0.72$ & $0.12$ & $0.08$ \\
BES4 B2 & $1.0$ & $10.0$ & $0.184$ & $1.35$ & $0.18$ & $0.09$ \\
BES4 C1 & $0.3$ & $1.0$ & $0.184$ & $0.22$ & $0.08$ & $0.07$ \\
BES4 C2 & $1.5$ & $1.0$ & $0.184$ & $0.22$ & $0.08$ & $0.07$ \\
MS A & $1.0$ & $1.0$ & $1.393$ & $0.73$ & $0.09$ & $0.05$ \\
MS B1 & $1.0$ & $5.0$ & $1.393$ & $2.39$ & $0.14$ & $0.08$ \\
MS B2 & $1.0$ & $10.0$ & $1.393$ & $4.49$ & $0.19$ & $0.09$ \\
MS C1 & $0.3$ & $1.0$ & $1.393$ & $0.58$ & $0.09$ & $0.06$ \\
MS C2 & $1.5$ & $1.0$ & $1.393$ & $0.85$ & $0.09$ & $0.06$ \\
\hline
\end{tabular}
\end{table*}

For all models considered, the value of $t_{amb}$ at the final density, $n_{\mathrm{H}} = 10^8$ cm$^{-3}$, is significantly lower than the collapse duration, while the { initial values are comparable to or larger than the collapse time, particularly for the BES4 collapse, and for the models with increased cosmic ray ionization rates}. At $10^4$ cm$^{-3}$, { the B1 and B2 models have $t_{\rm amb}$ larger than the collapse time for the BES4 case}. Increasing $\zeta$ leads to larger values of $t_{amb}$, as the ionization, and therefore the coupling to the neutral gas, is increased. Metallicity has very little effect at { higher densities, but the initial $t_{\rm amb}$ varies with the metallicity, as more readily-ionized atoms such as carbon are present}. The BES1 and MS models have { lower initial values of $t_{amb}$/$t_{\rm collapse}$} than the BES4 models, suggesting that the faster collapse should be impeded more strongly by the coupling of gas to magnetic fields. These results emphasize that ambipolar diffusion is important for diffuse material, where magnetic fields are likely to impede collapse, but once denser clumps have formed magnetic support will be removed too rapidly to affect the subsequent evolution.

\section{Discussion \& Conclusions}
\label{sec:discussion}

The results of our chemical modelling show that it is difficult to disentangle the effect on molecular abundances of different collapse modes from that of varying the input parameters, which are not necessarily known a priori. Drawing information about the dynamics of star formation from molecular abundances therefore requires a full investigation of parameter space, something which would be extremely time-consuming using combined hydrodynamical-chemical modelling. Our grid of models, although not large enough to draw robust conclusions about individual objects, does allow us to compare results with the general properties of prestellar cores, and determine whether particular collapse models or regions of parameter space are in conflict with observation.

{ Assuming the values from model A (see Table~\ref{tab:grid}), all density approximations predict CO abundances away from the core centre in agreement with observed values of $10^{-5}-10^{-4}$ in starless cores (\citet{caselli1999,frau2012}, assuming \(^{18}\mathrm{O}/^{16}\mathrm{O} \approx 10^{-3}\) and \(^{17}\mathrm{O}/^{16}\mathrm{O} \approx 10^{-4}\)). However, the AD collapse only reaches these values at $r \gtrsim 0.3 \pc$, much larger than typical core sizes ($< 0.1 \pc$; \citealt{frau2012}). Only models with the highest desorption efficiencies investigated for H$_2$ formation and cosmic-ray induced photodesorption (D2 and F2) predict CO abundances of $\sim 10^{-5}$ at a radius of $0.1 \pc$ for an ambipolar diffusion collapse.}

{ The BES1, BES4 and MS approximations result in similar abundance profiles for most molecules. The BES4 and MS abundances are generally more depleted in the centre than the BES1 ones, as the gas densities are higher due to either magnetic support, or higher initial densities and a more rapid collapse, causing more efficient freeze out. However, these differences are not large enough to provide a robust observational test of the mode of collapse. The AD approximation produces significantly different profile shapes to the other three, due to both the longer collapse duration and the higher densities at large radii. The slower increase with radius of the abundance of molecules such as CO, HCN and HCO$^+$, and the subsequent slow or negligible decline beyond the peak value, are qualitatively different to the situation with the initially unstable collapse modes, suggesting spatially resolved molecular observations could be used to discriminate between them.}

{ All density approximations predict peak NH$_3$ abundances of $\sim 10^{-8}$, consistent with observed values in prestellar cores \citep{tafalla2002,johnstone2010}. At $n_0 = 2 \times 10^{5} \pcc$, N$_2$H$^+$ abundances, { shown in Figure \ref{fig:n2h+},} are intermediate between the values found by \citet{frau2012} ($\sim 10^{-11}$) and \citet{tafalla2002} and \citet{johnstone2010} ($\sim 10^{-10}$). However, as with CO, for the AD approximation the abundances within the reported radii of the cores are much lower than the observed values. Ambipolar diffusion simulations including chemistry by \citet{tassis2012} show similar behaviour, with the inclusion of magnetic effects leading to a more extended depleted region than in unmagnetised models, suggesting that this is a genuine feature of collapse under ambipolar diffusion, rather than being due to our approximation of the density evolution.}

{ \citet{lee2004} calculated the chemical evolution of a quasistatically contracting BE sphere over $10^6 \yr$, finding similar abundance profiles to our BES1 approximation for CO, NH$_3$ and HCO$^+$, although for N$_2$H$^+$ we find much more depletion in the core centres, and higher HCN abundances overall. Given that our MS and BES4 approximations also give similar results to BES1, this suggests that the collapse timescale, rather than the specific details of the density evolution, are more important for the chemical evolution, at least for some observationally important molecules such as CO.}

{ Our BES1 and BES4 approximations are based on hydrodynamical simulations presented in \citet{aikawa2005}, who modelled the chemical evolution of these models self-consistently, providing a test of the approximations' accuracy. Comparing to their results, we find good agreement (of the same order of magnitude) between the predicted peak molecular abundances of both approaches. However, the variation with radius differs - our approximations generally predict lower abundances at the core centres, especially for the BES4 collapse, where our results predict lower CO abundances to the BES1 case at the same $r$, whereas \citet{aikawa2005} find the opposite. We attribute this to the differing initial conditions - \citet{aikawa2005} assume the gas is entirely atomic prior to collapse, whereas we allow the abundances to evolve for $10^6 \yr$ at the initial density before the onset of collapse, leading to significant freeze-out onto grains occuring in the denser central regions.}

{ We conclude that despite the dependence of molecular abundances on the various parameters mentioned above, molecular observations can still be useful for discriminating between different models of collapse. While models which begin from an initially gravitationally unstable state predict similar abundances and radial variations, ambipolar diffusion produces qualitatively different abundance profiles for many observationally important molecules, which appear to be in conflict with observations of prestellar cores, although varying the input parameters may be able to reduce this discrepancy.} A more exhaustive investigation of parameter space, combined with observations of multiple species from the same source, could be used to draw much stronger conclusions on the nature of core collapse, as well as providing constraints on the values of the input parameters which are currently assumed ad hoc. This sort of investigation would be extremely time consuming, if not impossible, using a coupled hydrodynamical-chemical system, even without the additional complications of magnetic fields. The results we have presented here demonstrate that these large grids of models are now feasible using our method of parametrizing the dynamics, while still providing a reasonable level of accuracy compared to full simulations.

\begin{figure}
  \plotone{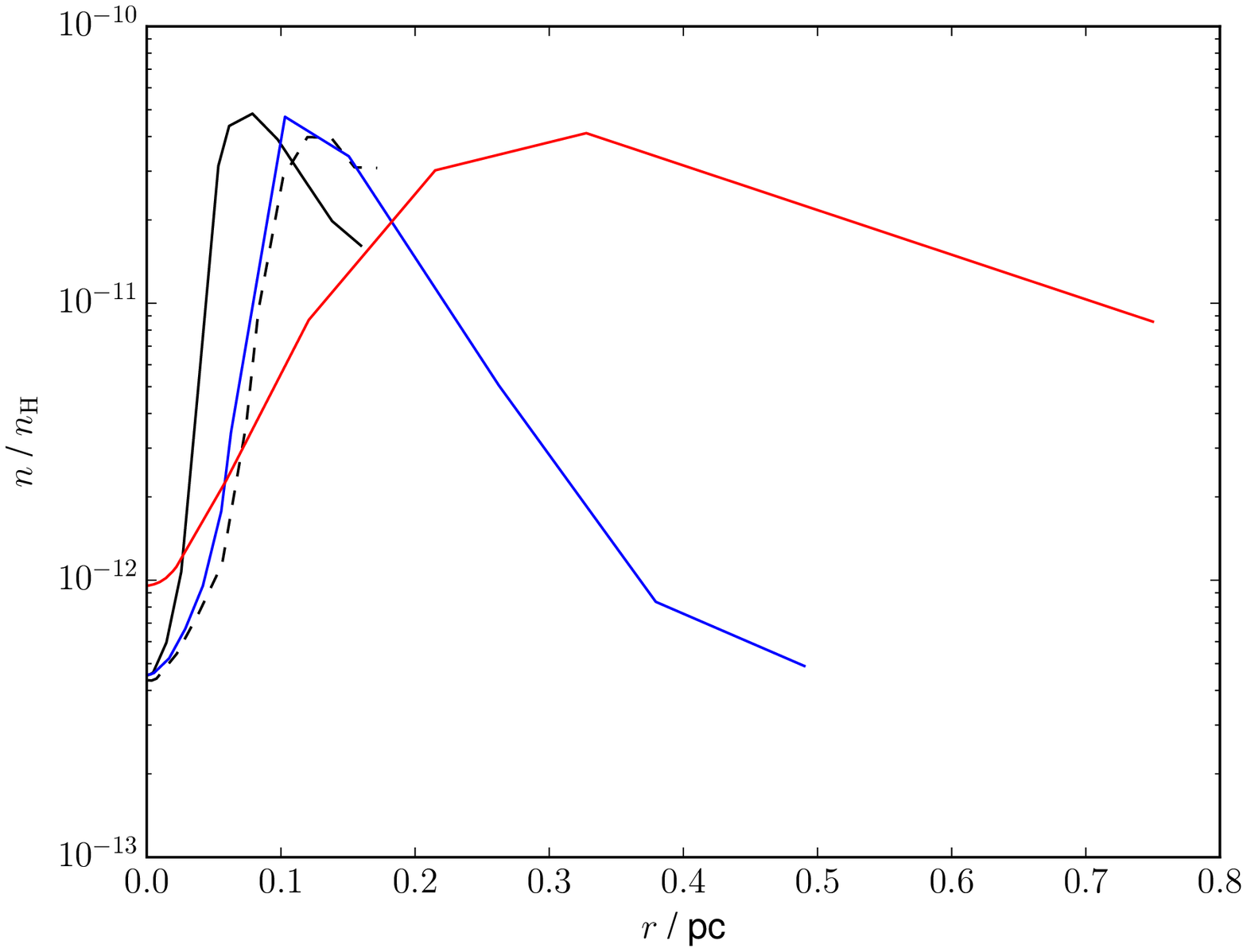}
  \caption{N$_2$H$^+$ abundance at a central density $n_0 = 2 \times 10^{5} \pcc$ for model A, using the BES1 (solid black), BES4 (dashed black), MS (blue) and AD (red) density approximations.}
  \label{fig:n2h+}
\end{figure}

\acknowledgments

FP is supported by the Perren fund and the IMPACT fund. We would like to thank Prof. Fred Adams for his helpful contribution, { and the referee, whose comments helped to produce a greatly improved version of the original paper}.

\software{UCL\_CHEM \citep{viti2004,holdship2017}}

%




\appendix

\section{Density approximations}
\label{sec:eqs}

The BES1, BES4 and AD collapse density profiles were approximated with the function
\begin{equation}
n(r) = \frac{n_0(t)}{1 + (\frac{r}{r_0(t)})^{a(t)}}
\end{equation}
whereas the MS collapse required a different functional form,
\begin{equation}
\rho(r) = \rho_0(t) \left(1 + \left(\frac{r}{r_0(t)}\right)^2\right)^{-a(t)}
\end{equation}
where $n_0(t)$, $\rho_0(t)$, $r_0(t)$ and $a(t)$ are functions of time since the onset of collapse given in the following subsections.

\subsection{BES1}

The time-dependent parameters are given by
\begin{equation}
  \log_{10} n_0(t) = 61.8 \left(1.175 \times 10^6 - t\right)^{-0.01} - 49.4
\end{equation}
\begin{equation}
  \log_{10} r_0(t) = -28.5 \left(1.175 \times 10^6 - t\right)^{-0.01} + 28.93
\end{equation}
\begin{equation}
  a(t) = 2.4
\end{equation}
where $n_0$ is in $\pcc$, $r_0$ is in AU and $t$ is in years.

\subsection{BES4}

The time-dependent parameters are given by
\begin{equation}
  \log_{10} n_0(t) = 68.4 \left(1.855 \times 10^5 - t\right)^{-0.01} - 55.7
\end{equation}
\begin{equation}
  \log_{10} r_0(t) = -39.0 \left(1.855 \times 10^5 - t\right)^{-0.01} + 38.7
\end{equation}
\begin{equation}
  a(t) = 1.9 + 0.5 \exp(-t / 10^5)
\end{equation}
where $n_0$ is in $\pcc$, $r_0$ is in AU and $t$ is in years.

\subsection{MS}

The time-dependent parameters are given by
\begin{equation}
  \log_{10} \rho_0(t) = 3.54 \left(5.47 - t\right)^{-0.15} - 2.73
\end{equation}
\begin{equation}
  \log_{10} r_0(t) = -1.34 \left(5.47 - t\right)^{-0.15} + 1.47
\end{equation}
\begin{equation}
  a(t) = 2.0 - 0.5\left(\frac{t}{5.47}\right)^9
\end{equation}
with the units determined by the initial central density $\rho_c$ and the sound speed, $c_s$. $\rho_0$ is in units of $\rho_c$, $r_0$ in units of $\frac{c_s}{\sqrt{2\pi G \rho_c}}$ and $t$ in units of $\left(2 \pi G \rho_c\right)^{-0.5}$.

\subsection{AD}

The time-dependent parameters are given by
\begin{eqnarray}
  \log_{10} n_0(t) = & \log_{10} (2 + 1.7 \left(\frac{t}{6}-1\right)) + 3 & \quad t < 6.0\\ 
                   & 5.3 \left(16.138 - t\right)^{-0.1} - 1.0 & \quad t \ge 6.0
\end{eqnarray}
\begin{equation}
  \log_{10} r_0(t) = -2.57 \left(16.138 - t\right)^{-0.1} + 1.85
\end{equation}
\begin{equation}
  a(t) = 2.4 - 0.2 \left(\frac{t}{16.138}\right)^{40}
\end{equation}
where $n_0$ is in $\pcc$, $r_0$ is in $0.75 \pc$ and $t$ is in Myr.

\section{Velocity profiles}
\label{sec:vr}

For the MS and AD collapses, the radial velocity profiles were also approximated in order to determine the inwards movement of the gas parcels.

\subsection{MS}

The radial velocity profile is given by
\begin{equation}
  v_r(r) = v_{\rm min}(t)\left[ \left(\frac{r'(t)}{r_{\rm min}(t)}\right)^2 - 1 \right]
\end{equation}
for $r < r_{\rm min}$ and
\begin{equation}
  v_r(r) = v_{\rm min}(t)\left[ \exp(-2a(t)r'(t)) - 2 \exp(-a(t)r'(t)) \right]
\end{equation}
for $r \ge r_{\rm min}$, where $r'(t) = r - r_{\rm min}$. The time evolution is given by
\begin{eqnarray}
                 & -1.149t + 7.2 & \quad t < 4.95 \\
  r_{\rm min}(t) = & -9.2 \log t + 16.25 & \quad t < 5.33 \\
                 & -22 \log t + 37.65 & \quad t \ge 5.33 
\end{eqnarray}
\begin{eqnarray}
                 & 0.0891 t & \quad t < 4.95 \\
  v_{\rm min}(t) = & 5.5 \log t -8.37 & \quad t < 5.33 \\
                 & 18.9 \log t -30.8 & \quad t \ge 5.33 
\end{eqnarray}
\begin{eqnarray}
                 & 0.0101 t + 0.4 & \quad t < 4.95 \\
  a(t) = & 0.695 \log t - 0.663  & \quad t < 5.33 \\
                 & 2.69 \log t - 4 & \quad t \ge 5.33 
\end{eqnarray}
where $v_{\rm min}$ is in units of $c_s$, $r_{\rm min}$ in units of $\frac{c_s}{\sqrt{2\pi G \rho_c}}$ and $t$ in units of $\left(2 \pi G \rho_c\right)^{-0.5}$.

\subsection{AD}

The radial velocity profile is given by
\begin{eqnarray}
           & v_{\rm min}(t)\left[ \left(\frac{r'(t)}{r_{\rm min}(t)}\right)^2 - 1 \right] & \quad r < r_{\rm min} \\
  v_r(r) = & (v_{\rm min}(t) - v_{\rm mid}(t)) \left( \frac{r'(t)}{0.5 - r_{\rm min}(t)} \right)^{0.3} - v_{\rm min}(t) & \quad r < 0.5 \\  
           & 2v_{\rm mid}(t) (r - 1) & \quad r \ge 0.5
\end{eqnarray}
The time evolution is given by
\begin{eqnarray}
                 & -0.0039t + 0.49 & \quad t \le 10.2 \\
  r_{\rm min}(t) = & -0.0306(t-10.2) + 0.45 & \quad t \le 15.1 \\
                 & -0.282(t-15.1) + 0.3 & \quad t > 15.1
\end{eqnarray}
\begin{eqnarray}
  v_{\rm min}(t) = & 3.44(16.138-t)^{-0.35} - 0.7
\end{eqnarray}
\begin{eqnarray}
  v_{\rm mid}(t) = & 0.143t & \quad t \le 10.2 \\
                 & 0.217(t-10.2) + 1.46 & \quad t > 10.2
\end{eqnarray}
where $r_{\rm min}$ is in units of $0.75 \pc$, $v_{\rm min}$ and $v_{\rm mid}$ are in $10^{-2} \, {\rm km} \, {\rm s}^{-1}$ and $t$ is in Myr.




\bibliographystyle{aasjournal}
\bibliography{collapse}



\end{document}